\documentclass[twocolumn,times]{aastex62}
\usepackage{amsmath}
\usepackage{amssymb}

\usepackage{bm}
\usepackage{color}
\allowdisplaybreaks[4]

\def\vv#1{\bm{#1}}
\def\be {\begin{equation}}
\def\ee {\end{equation}}
\def\vR {\vv{R}}
\def\ddvR{\vv{\ddot R}}

\def\vrkl {\vv{r}_{kl}}

\def\urkl {\vv{\hat r}_{kl}}
\def\rkl {r_{kl}}

\def\vF{\vv{F}}
\def\vT {\vv{T}}

\def\ddvR{\vv{\ddot R}}

\def\dvL{\vv{\dot L}}

\def\TI {{\bf \cal I}}

\def\TJ {{\bf \cal J}}

\def\m {m}

\def\tt {\{t\}}

\usepackage{color}

\shorttitle{Origin of the Moon}
\shortauthors{Wenshuai Liu}

\begin{document}

\title{\large{\textbf{Origin of the lunar inclination from tidal interaction of multiple-moon system}}}

\correspondingauthor{Wenshuai Liu}
\email{674602871@qq.com}

\author{Wenshuai Liu}
\affiliation{School of Physics, Henan Normal University, Xinxiang 453007, China}



\begin{abstract}
According to the giant impact theory, the Moon formed through accreting the debris disk produced by a collision between Theia and the proto-Earth, and the predicted lunar orbital inclination relative to the Earth's equatorial plane is about within one degree when Moon formed. However, the current lunar orbital inclination with five degrees relative to the Earth's orbital plane requires the Moon's orbital inclination relative to the Earth's equator to be about ten degrees when traced back to the time of lunar formation. Since two moons are also a natural outcome of simulations of lunar formation from a protolunar disk produced by a giant impact, here we show that, under solar perturbation, gravitational tidal interaction between Earth and its two moons with negligible orbital inclination relative to Earth's equatorial plane could lead to a merger of one moon with Earth, or a merger of the two moons or an ejection of one moon, resulting that the surviving moon's orbital inclination relative to Earth's equator could exceed ten degrees. The theory proposed here may provide a way of explaining the initial large lunar inclination relative to the Earth's equator.
\end{abstract}



\section{Introduction}

The giant impact theory is the leading explanation for the formation of the Moon, positing that the Moon formed from debris resulting from a collision between a planet-sized body and the proto-Earth \citep{1,2}. This theory predicts that following lunar accretion, the Moon would achieve a near-equatorial orbit, with an initial lunar inclination of approximately within one degree relative to the Earth's equatorial plane \citep{3}. This is attributed to the circum-terrestrial disk's tendency to settle within the Earth's equatorial plane on a timescale significantly shorter than that of its evolutionary timescale \citep{3,4}. As suggested by tidal evolution calculations that for every degree of lunar orbital inclination relative to the Earth's equatorial plane when the Earth-Moon separation is about 10 Earth radii (10$R_E$) \citep{5,6,7}, the angle between the present lunar orbital plane and the Earth's orbital plane could be approximately half a degree, however, current observations indicate that the lunar orbital inclination relative to the Earth's orbital plane is approximately five degrees, suggesting that the Moon's orbital inclination relative to the Earth's equatorial plane at a distance of 10$R_E$ would have been around 10 degrees \citep{4}. This observation presents a contradiction to the results derived from forward integrations of the lunar orbit with an initial lunar inclination of approximately one degree \citep{3}. The contradiction between the theoretical predictions regarding the initial lunar inclination and the values inferred from current observations gives rise to what is known as the lunar inclination problem, an issue that remains in debate in the field of lunar studies.

Theories proposed to account for this puzzle include Moon-Sun resonances \citep{8}, interactions between the newly formed Moon and the protolunar disk \citep{9}, collisionless encounters with large planetesimals \citep{4}, and interactions with a high initial obliquity Earth \citep{10,11}. As discussed in \citep{4}, neither of the first two scenarios is satisfactory. The third scenario also faces difficulty \citep{10}. The last theory needs an initial obliquity $\theta>61^\circ$ of the Earth with a much faster spin. In this study, we introduce a mechanism by which the initial lunar inclination may originate from a multiple-moon system formed after the giant impact. According to this theory, two moons emerged from the debris disk following the giant impact. Influenced by solar perturbations, the gravitational interactions, coupled with tidal effects between the two moons and Earth, could lead to the merger of one moon with Earth, the coalescence of the two moons, or the ejection of one moon. Such dynamics could consequently allow the remaining moon's initial orbital inclination relative to Earth's equatorial plane to reach values larger than 10 degrees. The Earth's spin period within the range predicted by canonical giant impact is adopted in this work. In order to use such spin period, lunar isotopic crisis should be explained in the framework of the canonical giant impact theory.

Theory accounting for Earth-Moon isotopic similarity in the framework of the canonical giant impact theory is discussed in Section 2. Lunar inclination from tidal interaction of Earth with its two moons is in Section 3. The discussions are in Section 4.

\section{Earth-Moon isotopic similarity}
The tidal interactions between Earth and its two moons are significantly influenced by Earth's rotation set by the giant impact. The canonical giant impact hypothesis \citep{15} posits that the post-impact angular momentum of the Earth-Moon system aligns closely with its present-day state. However, numerical simulations based on this hypothesis indicate that the circum-terrestrial debris disk, from which the Moon formed, consists of a substantial proportion of materials from the impactor \citep{15,16,17}. Considering that various bodies within the Solar System possess distinct compositional characteristics, one would expect the composition of the Moon, as inferred from the giant impact model, to differ notably from that of Earth. In contrast, isotopic analyses of samples from both Earth and the Moon demonstrate a remarkable similarity in their isotopic signatures \citep{18,19,20,21,22,23}, giving rise to the phenomenon known as the lunar isotopic crisis. To explain the observed isotopic similarity between Earth and its Moon, it has been suggested that the materials constituting the Moon either equilibrated with or were derived from the Earth's mantle subsequent to the impact \citep{24,22}, or that the Moon originated from a giant impact with an impactor with isotopic characteristics identical to those of Earth \citep{26}.

With the constraint of modern angular momentum of the Earth-Moon system, \cite{26} investigated the condition of an Earth-like composition impactor through N-body simulations of the late stages of accretion, showing the impactor and Earth could originate from a common source. However, N-body simulations conducted by \cite{37} indicate that the likelihood of the proto-Earth and the impactor exhibiting similar isotopic ratios is less than $5\%$. \cite{38} suggested that the orbiting disc and the hot, post-impact Earth can effectively exchange material through a vapor cloud that envelops both the molten disc and the planet, but analyses in \cite{39} showed that to allow substantial material exchange and isotopic compositional equilibrium, the amount of exchanged material must be about 4.6 times the total mass of the accretion disk, which would inevitably involve considerable angular momentum transfer between the early Earth and the accretion disk and could disrupt the orbital stability of the accretion disk.

Without the constraint of modern angular momentum of the Earth-Moon system, \cite{27} proposed a modified version of the giant impact theory that incorporates a rapidly rotating proto-Earth. In this model, a greater portion of material from proto-Earth could be delivered to the circum-terrestrial debris disk, which is predominantly composed of material from the proto-Earth and thereby accounts for the Earth-like isotopic composition of the Moon. This disk is sufficiently massive to potentially form the Moon following the collision of a body slightly smaller than Mars with the fast-spinning proto-Earth. This hypothesis has the potential to explain the observed isotopic composition of the Moon, but it results in an angular momentum that exceeds that of the contemporary Earth-Moon system. In addition, the initial fast spin of the Earth required in \cite{27} is challenged by a sequence of multiple impacts \citep{41,42} due to the fact that an initial fast spin of the Earth is disfavored by angular momentum drain by multiple impacts. Furthermore, multiple impacts model proposed in \cite{41} may relax isotopic and angular momentum constraints since the moonlet produced by each impact could accrete onto the early Moon. Other models characterized by high energy and high angular momentum, including Synestia \citep{28,29} and the collision of two bodies of comparable mass \citep{30}, similarly yield substantial angular momentum that must be reduced through a viable mechanism that remains in debate.

In this work, we set the initial spin of Earth to be 4.8 hours which is within the range predicted by the canonical giant impact.

\section{Tidal interaction of Earth with its two moons under solar perturbation}

Previous studies \citep{3} have demonstrated that two large moons can form within a year and are a common outcome of giant impacts. Therefore, it is essential to explore the tidal evolution of interactions among the Earth and its two moons under solar perturbation, particularly when the two moons' initial orbital planes are not aligned with the Earth's equator. \cite{12} examined the coplanar scenario without considering solar perturbation. In this study, we investigate the non-coplanar configuration with solar perturbation to investigate whether this configuration can provide answers to the puzzle of lunar inclination.

\cite{3} showed that the formed single moon's initial orbital inclination to the Earth's equator is about within $\mathrm{1}^\circ$. We assume each moon's initial orbital inclination to the Earth's equator is $\mathrm{1}^\circ$ in the condition of Earth-moon-moon configuration. According to the results in \cite{3}, the masses of the two moons are set to be $\mathrm{M_{1}=0.39M_L}$ (the low mass moon) and $\mathrm{M_{2}=0.63M_L}$ (the large mass moon) where $\mathrm{M_L}$ is the present lunar mass. The orbits of the two moons are nearly circular with eccentricity $e_1=e_2=0.0001$ and with semimajor axis $\mathrm{a_1=0.93a_R}$ and $\mathrm{a_2=1.98a_R}$ where $\mathrm{a_R}$ is the Roche limit of Earth and $\mathrm{a_R}\approx 2.9R_E$ ($R_E$ is the radius of the Earth). The separation between the Sun and Earth is $\mathrm{1AU}$ with Earth's eccentricity to be zero and obliquity to be $\mathrm{10^\circ}$. It should be noted that the moon with $\mathrm{a_1=0.93a_R}$ shown in \cite{3} is within the Roche limit and would be disrupted by Earth's tidal force. The reason for the presence of moon within the Roche limit is that simulations in \cite{3} only considered merger of particles and moon which formed outside the Roche limit would not be allowed to be disrupted when moving into the Roche limit later on. Here, for simplicity, we adopt $\mathrm{a_1=0.93a_R}$ but change the density of the moons in order that the surface of the low mass moon with the new density is tangent to the new calculated Roche limit based on the new density of the moons (just beyond the Roche limit). The resulting new density of the moons is $\mathrm{5087g/cm^3}$, and the radius of the low mass moon and the large mass moon are 1100km and 1300km, respectively.

\begin{figure*}
     \begin{tabular}{cc}
            \includegraphics[width=0.5\textwidth]{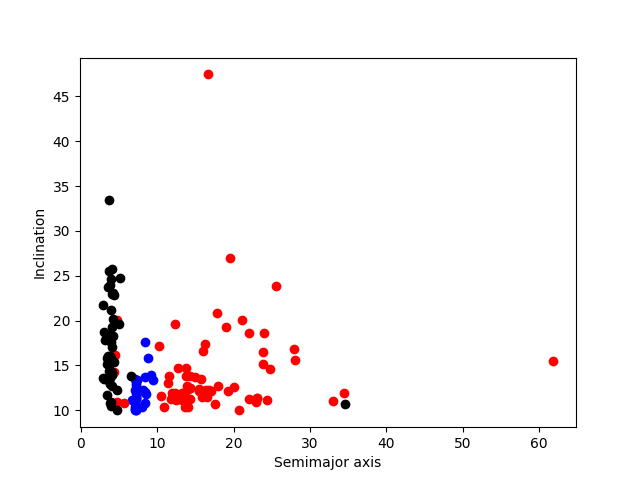}
            \includegraphics[width=0.5\textwidth]{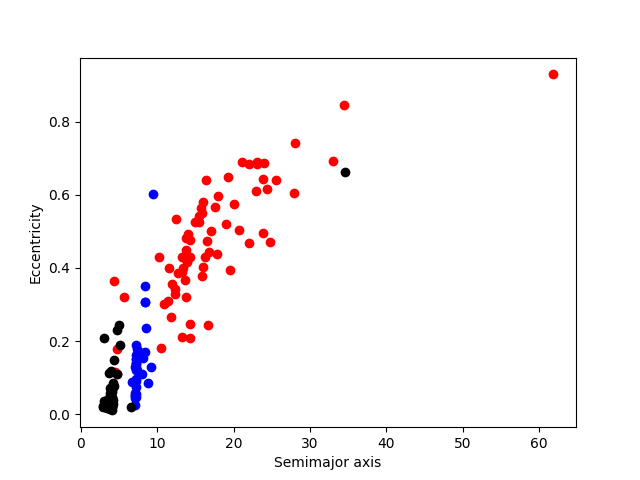}
            \end{tabular}
\caption{Left and right show the surviving moon's orbital inclination relative to the Earth's equatorial plane and eccentricity along with semimajor axis from the 156 runs. Black dots, blue dots and red dots represent the outcomes of the survival of the low mass moon, the merger of the two moons and the survival of the large mass moon, respectively. Semimajor axis is in unit of Earth radius and inclination is in unit of degree.}
\label{fig:figure2}
\end{figure*}

\begin{figure*}
     \begin{tabular}{cc}
            \includegraphics[width=0.225\textwidth]{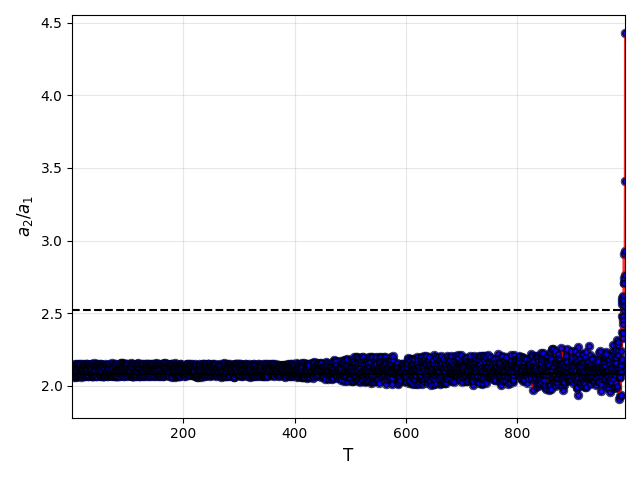}
            \includegraphics[width=0.25\textwidth]{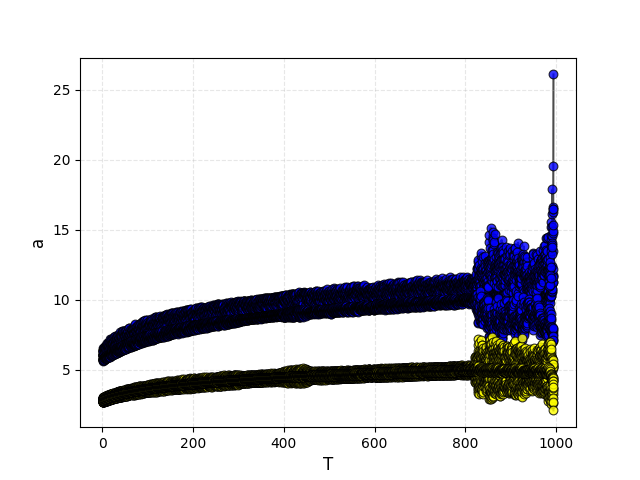}
            \includegraphics[width=0.25\textwidth]{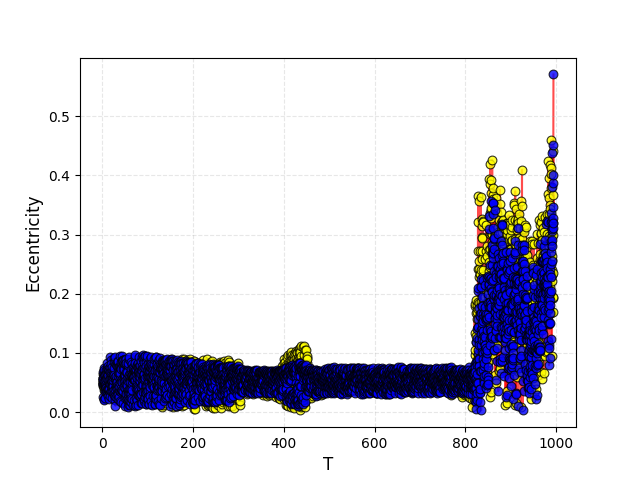}
            \includegraphics[width=0.25\textwidth]{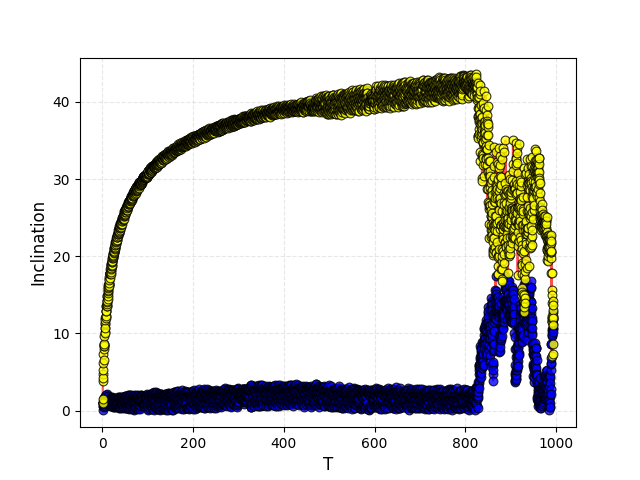}\\
            \includegraphics[width=0.225\textwidth]{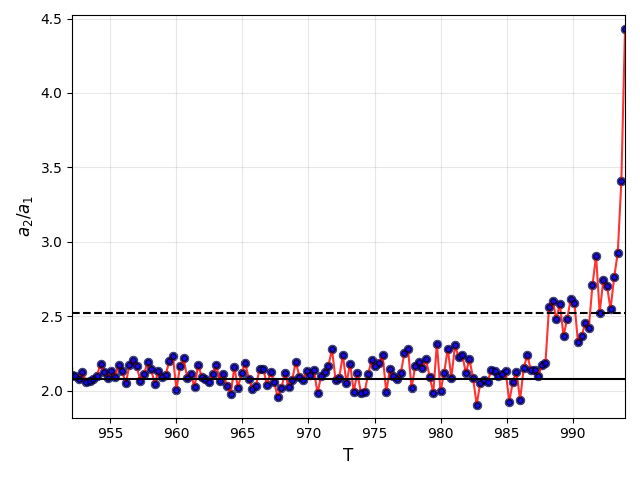}
            \includegraphics[width=0.25\textwidth]{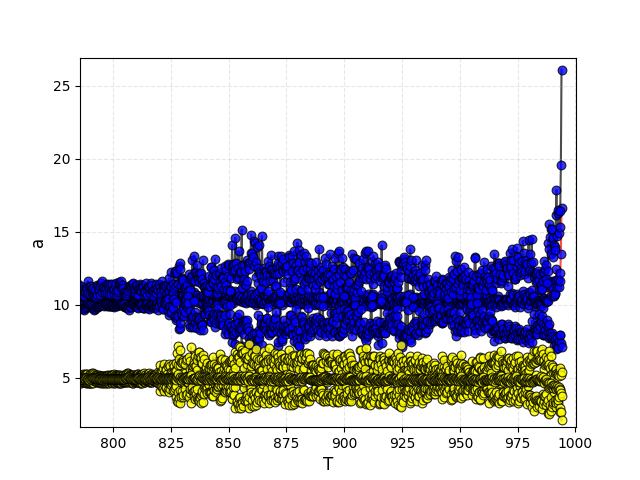}
            \includegraphics[width=0.25\textwidth]{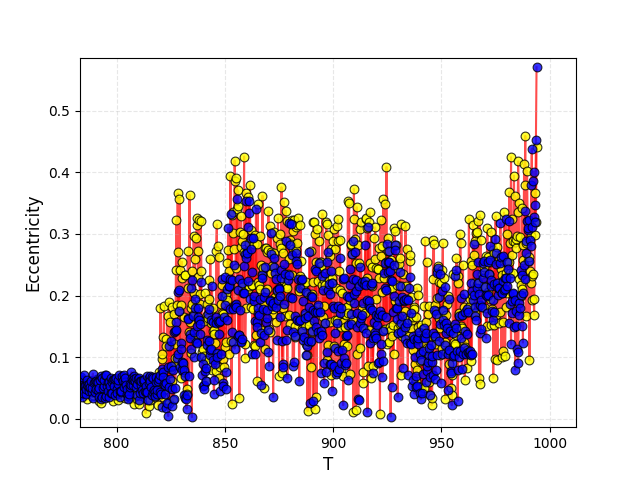}
            \includegraphics[width=0.25\textwidth]{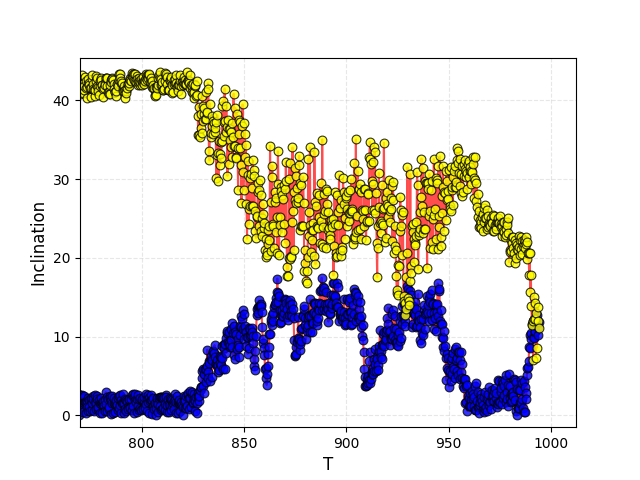}
            \end{tabular}
\caption{The first column is the ratio the semimajor axis of the large mass moon to that of the low mass moon at each timestep, and the black line and the black dotted line represent $3:1$ MMR and $4:1$ MMR, respectively. The second column presents the semimajor axis, periapse and apoapse of the two moons at each timestep, blue dots and yellow dots represent the results of the large mass moon and that of the low mass moon, respectively. The third and fourth column are the eccentricity and the inclination at each timestep, respectively. The blue dots and yellow dots in the third and fourth column represent the results of the large mass moon and that of the low mass moon, respectively. The lower panel is a close-in view of the upper panel. T is in unit of year. The semimajor axis, periapse and apoapse of the two moons are in unit of Earth radius, and inclination is in unit of degree.}
\label{fig:figure2}
\end{figure*}

\begin{figure*}
     \begin{tabular}{cc}
            \includegraphics[width=0.225\textwidth]{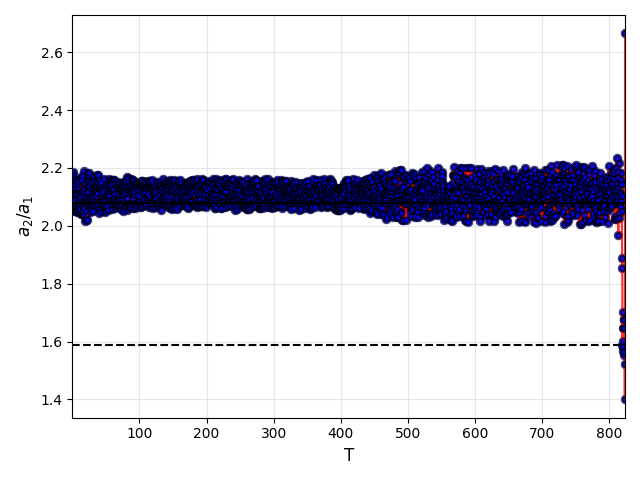}
            \includegraphics[width=0.25\textwidth]{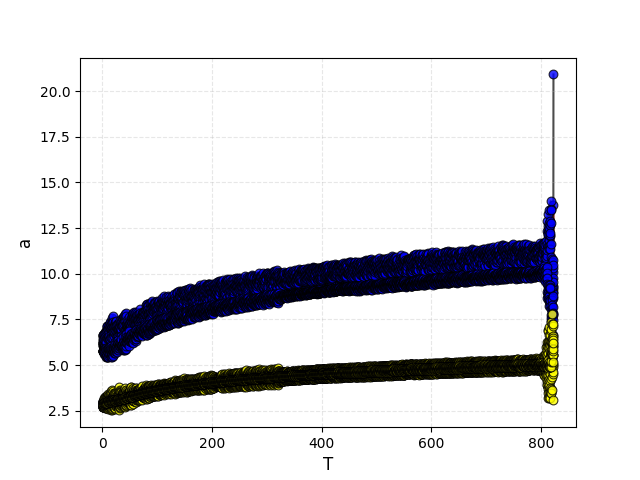}
            \includegraphics[width=0.25\textwidth]{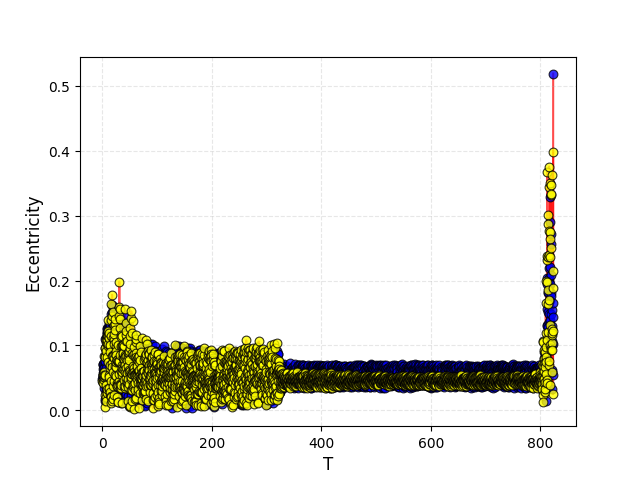}
            \includegraphics[width=0.25\textwidth]{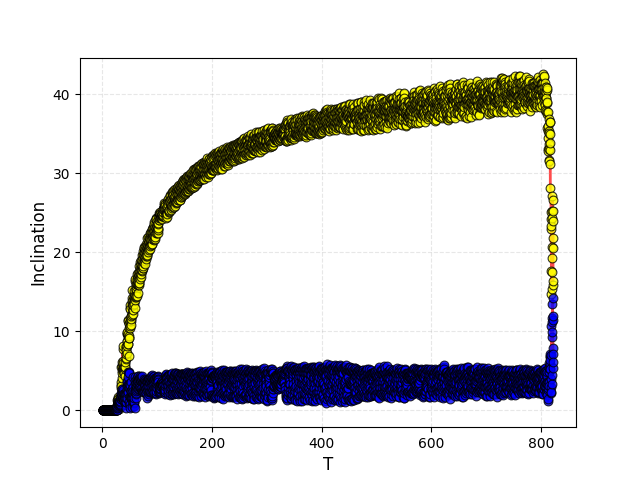}\\
            \includegraphics[width=0.225\textwidth]{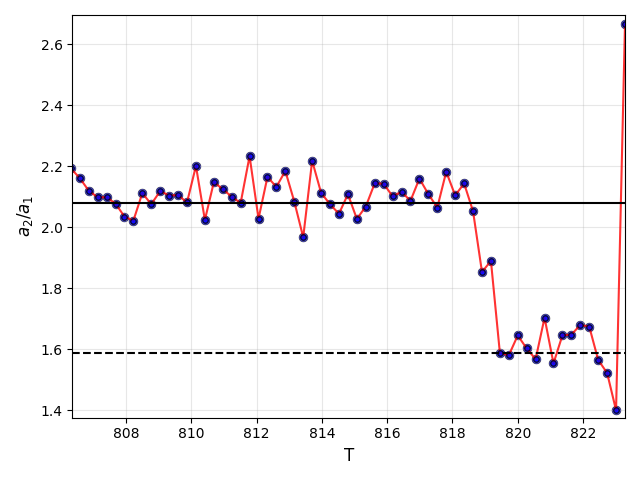}
            \includegraphics[width=0.25\textwidth]{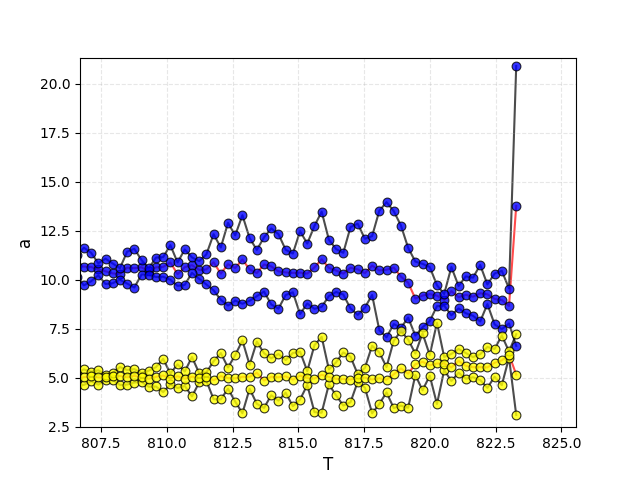}
            \includegraphics[width=0.25\textwidth]{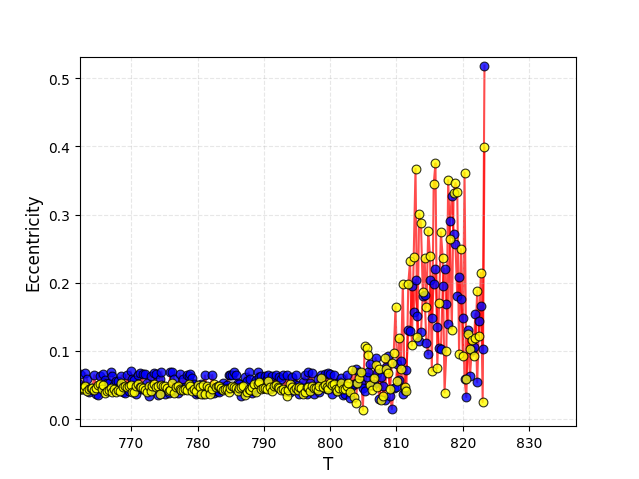}
            \includegraphics[width=0.25\textwidth]{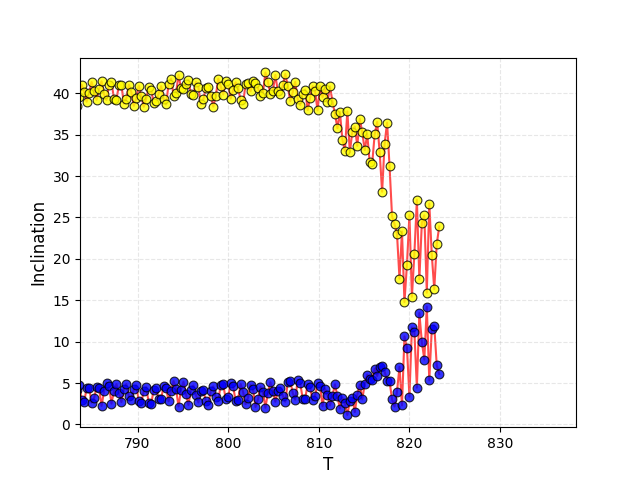}
            \end{tabular}
\caption{Same as Figure 2 but the two moons' initial inclinations relative to Earth's equator are $0^\circ$ and the large mass moon's initial longitude of ascending node is $\Omega_2=121^\circ$. The
black line and the black dotted line in the first column represent 3 : 1 MMR and 2 : 1 MMR, respectively.}
\label{fig:figure2}
\end{figure*}

\begin{figure*}
     \begin{tabular}{cc}
            \includegraphics[width=0.225\textwidth]{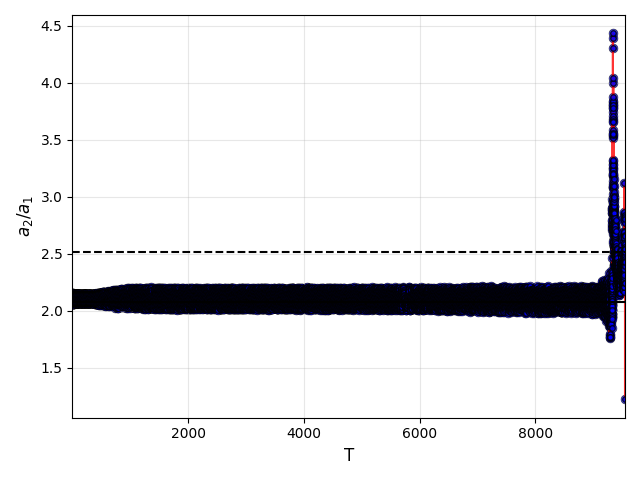}
            \includegraphics[width=0.25\textwidth]{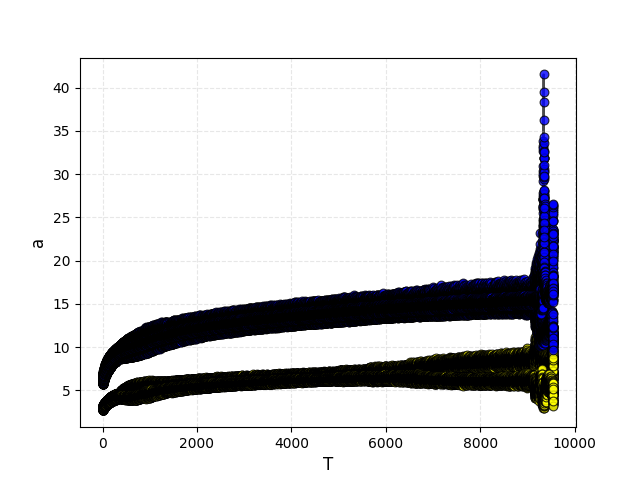}
            \includegraphics[width=0.25\textwidth]{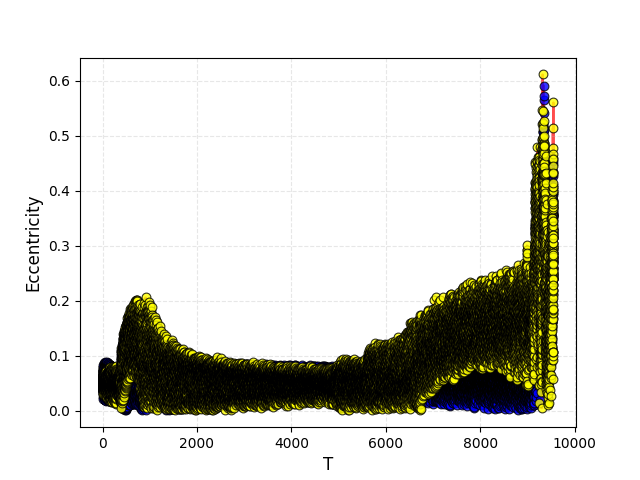}
            \includegraphics[width=0.25\textwidth]{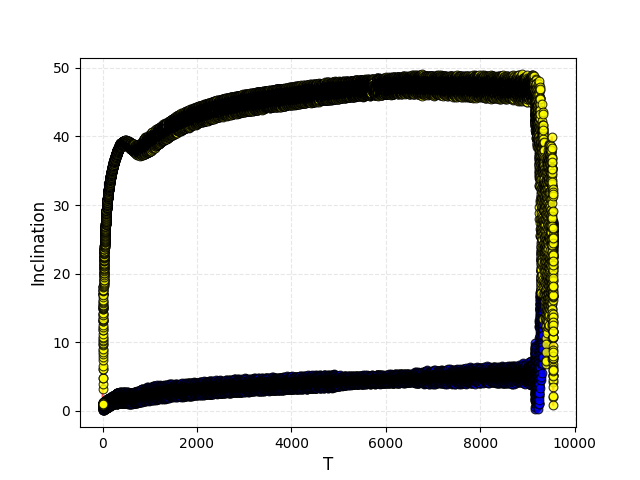}\\
            \includegraphics[width=0.225\textwidth]{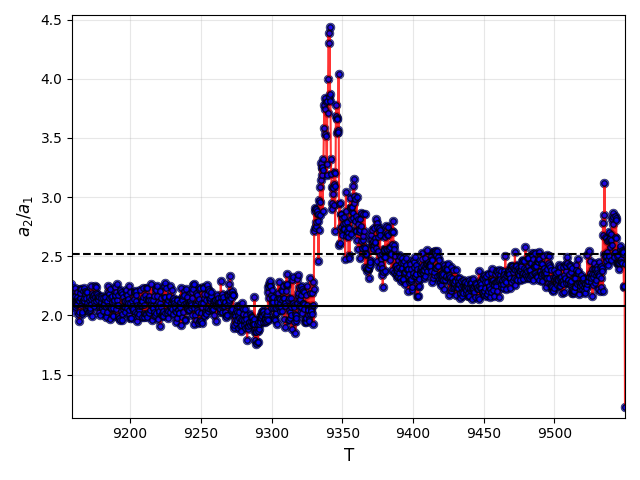}
            \includegraphics[width=0.25\textwidth]{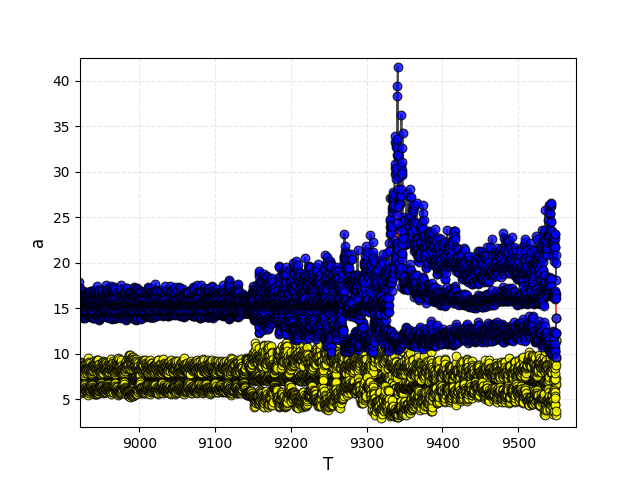}
            \includegraphics[width=0.25\textwidth]{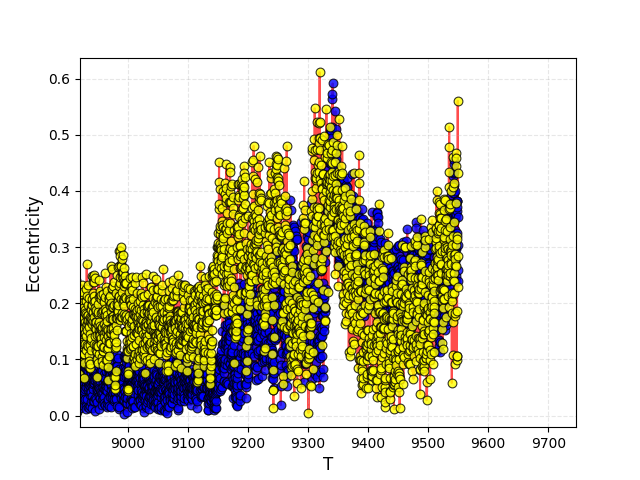}
            \includegraphics[width=0.25\textwidth]{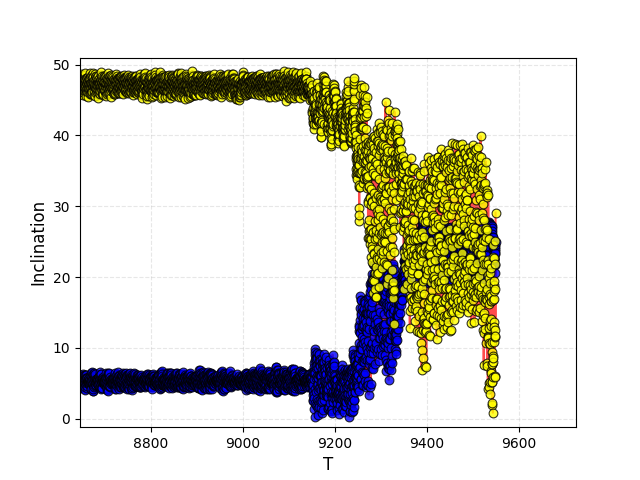}
            \end{tabular}
\caption{Same as Figure 2 but without solar perturbation and the large mass moon's initial longitude of ascending node is $\Omega_2=149^\circ$. The
black line and the black dotted line in the first column represent 3 : 1 MMR and 4 : 1 MMR, respectively.}
\label{fig:figure2}
\end{figure*}

To study the tidal evolution of Earth-moon-moon system under solar perturbation, we use code TIDYMESS \citep{13}, a N-body code with tidal force, to simulate the four-body problem with mutual tidal interaction. For N-body with tidal force, each with masses $\m_k$ and inertia tensors $\TI_k$ ($k = 1,...,N$), the total potential energy is shown as \citep{13}
\begin{equation}
U = \sum_{k=1}^N \sum_{l>k}^N \left( - \frac{G \m_k \m_l}{\rkl}
+ \frac{3 G}{2 \rkl^3} [ \urkl \cdot \TJ_{kl} \cdot \urkl - \frac{1}{3} \mathrm{tr}(\TJ_{kl}) ] \right)
\end{equation}
where $\TJ_{kl} = \m_k \TI_l + \m_l \TI_k$, and $\vrkl = \vR_k - \vR_l$. $\vR_k$ is the position of the center-of-mass of the body $k$ in the inertial frame.
Then, the equations of motion of body $k$ in the inertial frame are \citep{13}
\begin{equation}
\ddvR_k = \vv{a}_k = \frac{1}{\m_k} \sum_{l\ne k} \vF_{kl} (\vrkl)
\end{equation}
\begin{equation}
\dvL_k = \sum_{l\ne k} \vT_{kl} (\vrkl)
\end{equation}
where $\vv{L}_k$ is the rotational angular momentum vectors of the body $k$. $\vF_{kl} (\vrkl)$ and $\vT_{kl} (\vrkl)$ are given in details in \citep{13}.

Tidal interaction arises from tidal potential, which is usually derived based on a lag angle $\delta$. The angle $\delta$ has a relationship with the tidal time delay $\Delta t$ shown as $\delta=(\omega-n)\Delta t$ where $\omega$ is the angular rotation rate of Earth and $n$ is the moon's mean motion. Due to $\Delta t$ being much smaller in the early history of Moon, we set the tidal time delay to be $\Delta t_E=60s$ for Earth instead of about 12 minutes adopted in \cite{12} and $\Delta t_M=30s$ for moons. The fluid second Love number of Earth and the two moons adopted in TIDYMESS are $\mathrm{k_{2f,E}=0.993}$ and $\mathrm{k_{2f,M}=0.9}$, respectively. The adopted $\mathrm{k_{2f,E}=0.993}$ is same as that in \cite{33}. For the initial position of the two moons with both zero longitude of pericenter and zero true anomaly, we set the low mass moon's longitude of ascending node to be $\Omega_1=0^\circ$ and the large mass moon's longitude of ascending node $\Omega_2$ to be in the range of $0^\circ$ to $360^\circ$ with interval of $1^\circ$. For the initial rotation rate of Earth and the moons, we set the initial rotational period of Earth to be $\mathrm{4.8 \enspace hours}$ and assume that the two moons' initial spin periods are equal to their initial orbital periods.

Simulations of tidal evolution will be terminated if collision or ejection occurs. It should be noted that debris would be generated by collision of the moons and the amount of debris would be enhanced if collision occurs within Earth's gravitational potential, resulting that the mass retained in the final Moon becomes small \citep{44}. Here, for simplicity, collision which occurs if the distance between the two moons' center of mass is smaller than the sum of the radius of the two moons, is assumed to be a perfectly plastic merger as that in \cite{33}.

After tidal evolution of several hundred years simulated using TIDYMESS with Creep$\_$tides$\_$tidymess implementation \citep{13}, a moon can merge with the other moon or Earth, or be ejected, resulting that the surviving moon's orbital inclination relative to the Earth's equatorial plane can exceed $10^\circ$. As shown in Figure 1, 34 runs of the 360 simulations end with the merger of the two moons and the newly formed moon has an orbital inclination above $10^\circ$ relative to the Earth's equatorial plane, 71 runs of the 360 simulations end with the low mass moon colliding with Earth or being ejected and the surviving large mass moon's orbital inclination relative to the Earth's equator is above $10^\circ$. 51 runs of the 360 simulations end with the large mass moon colliding with Earth or being ejected and the surviving low mass moon has an orbital inclination above $10^\circ$ relative to the Earth's equator. It shows from Figure 1 that the semimajor axis of the surviving large mass moon is mainly above 10$\mathrm{R_E}$, the surviving low mass moon's semimajor axis is almost less than 5$\mathrm{R_E}$ and the semimajor axis of the newly formed moon resulting from the merger of the two moons falls into the range between 5$\mathrm{R_E}$ and 10$\mathrm{R_E}$.

Figure 1 indicates that the mechanism of the tidal interaction between Earth and its two moons under solar perturbation in the early time could account for the initial large lunar inclination but only shows the semimajor axis, eccentricity and inclination of the surviving moon. In reality, orbital evolution with tidal interaction can be quite complex, Figure 2 presents an example of a typical simulation that results in an ejection of the low mass moon. The large mass moon's initial longitude of ascending node is set to be $\Omega_2=61^\circ$, and the resulting semimajor axis, eccentricity and inclination of the surviving moon are 15.85$\mathrm{R_E}$, 0.55 and $11.8^\circ$, respectively. In Figure 2, it shows that the two moons encounter $\mathrm{3:1}$ mean motion resonance which gradually increases the inclination of the low mass moon in the fist 800 years while the large mass moon's inclination has no significant increase. Later on, two moons enter $\mathrm{4:1}$ mean motion resonance before and during which eccentricities of the two moons are excited significantly, resulting in close encounter of the two moons. In turn, close encounter can strongly affect the evolution of the two moons' inclination as shown in the fourth column of Figure 2.

The simulations in Figure 1 and Figure 2 are conducted with the initial condition that each moon's initial orbital inclination relative to the Earth's equator is $\mathrm{1}^\circ$. Does the simulation with each moon's initial orbital inclination equal to $\mathrm{0}^\circ$ and with solar perturbation also produce an initial large lunar inclination? In order to test this extreme initial condition, each moon's initial orbital inclination relative to the Earth's equator is set to be $\mathrm{0}^\circ$, the large mass moon's initial longitude of ascending node is $\Omega_2=121^\circ$ and other parameters are same as that adopted in Figure 2. The result of this simulation is that a merger of the two moons happens, and the semimajor axis, eccentricity and inclination of the newly formed moon are 7.5$\mathrm{R_E}$, 0.14 and $10.5^\circ$, respectively. The orbital evolution in this case is shown in Figure 3. The two moons are captured into $\mathrm{3:1}$ mean motion resonance and the inclination of the low mass moon is increased gradually in the fist 800 years while, same as that in Figure 2, the large mass moon's inclination is not excited significantly. After about 819 years, the two moons encounter $\mathrm{2:1}$ mean motion resonance which leads to the increase of the two moons' eccentricities and close encounter of the two moons. Close encounter of the two moons can have significant effect on the inclination of the two moons as shown in the fourth column of Figure 3.

So, why the extreme initial condition above could also lead to large lunar inclination? In order to answer this question, we conduct another simulation with the same parameters as that adopted in Figure 2 but with the large mass moon's initial longitude of ascending node $\Omega_2=149^\circ$ and without solar perturbation. The result is that the low mass moon is ejected, and the semimajor axis, eccentricity and inclination of the remaining moon are 12$\mathrm{R_E}$, 0.3 and $15^\circ$, respectively. The two moons' orbital evolution with respect to this case is shown in Figure 4. It indicates from this case that initial minor orbital inclination of the two moons could also produce large lunar inclination even without solar perturbation. From this point of view, effect of solar perturbation on the two moons' orbit with initial zero orbital inclinations adopted in Figure 3 could generate minor orbital inclination of the two moons' orbital planes relative to Earth's equatorial plane, then, as shown in Figure 4, such minor inclination would produce large lunar inclination through tidal interaction between Earth and its two moons.

\section{Discussion}
We propose a new mechanism for the origin of lunar inclination. Tidal interaction between Earth and its two moons under solar perturbation shows to be a viable way of producing initial high lunar inclination. Since two moons are also a natural outcome of simulations of lunar formation from a debris disk resulting from the giant impact, our mechanism has confirmed that, under solar perturbation, tidal interaction between Earth and two moons with initial negligible orbital inclination relative to Earth's equator could result that a moon merges with the other moon or Earth, or is ejected, and that the surviving moon's orbital inclination relative to Earth's equator could exceed $10^\circ$. Similar outcomes appear even in the extreme initial condition that the orbital plane of each moon is coplanar with Earth's equatorial plane under solar perturbation, meaning that lunar inclination problem may be a natural result of the tidal interaction between Earth and its two moons in the early time.

Although two moons mass of each is less than lunar mass is adopted in this work, it is expected that tidal interaction between Earth and two moons one of which has lunar mass would also result in a large inclination of the lunar mass moon after the other moon is ejected or merges with Earth. However, the detailed configuration of such two moons is the result of the moon-forming N-body simulation which is beyond the scope of this work. In addition, unusually high densities used for the moons are adopted in order to make the initial orbit of the low mass moon just beyond the Roche limit, meaning that some collisions may have been missed due to the resulting smaller radius of the moons. Thus, mergers may be more common under a more realistic setup.

The prediction that the Moon had an early inclination of 10 degrees as reconstructed by pioneering works \citep{5,6,7} is partially an artifact of not including lunar obliquity tides, which were first recognized as an important factor for the evolution of lunar inclination by \cite{43}. With long-term average tidal dissipation within Earth and current lunar tidal properties, lunar obliquity tides can greatly decrease lunar inclination during the Cassini state transition \citep{10}, indicating that lunar inclination may be higher before the Cassini state transition, possibly as high as 30 degrees when integrated back in time with lunar inclination of present value of 5 degrees (see Extended Data Fig. 1 in \cite{10}). However, lunar inclination during the Cassini state transition is strongly affected by lunar tidal quality factor $\mathrm{Q_M}$, the larger $\mathrm{Q_M}$ lead to less damping of lunar inclination. As shown in Extended Data Fig. 1 in \cite{10}, lunar tidal quality factor with current value $\mathrm{Q_M=38}$ would lead to lunar inclination of about 30 degrees before Cassini state transition but about 7 degrees if $\mathrm{Q_M=10000}$ is adopted. Similar results can also be found in \cite{45} where $\mathrm{Q_M}$ with small value leads to a large initial lunar inclination when integrated back in time. Since lunar tidal quality factor has evolved over time (like Earth's tidal quality factor, lunar tidal quality factor was significantly higher in the early time), the real lunar tidal quality factor along with time may lead to an initial lunar inclination with value which falls into the range predicted by the tidal evolution of multiple-moon system proposed in this work.


\section*{ACKNOWLEDGEMENTS}
We are grateful for the code TIDYMESS in \cite{13}. This work is supported by Natural Science Foundation of Henan (252300420327).

\end{document}